# On Near-Earth Asteroid Study at Department of Astronomy, Bandung Institute of Technology

S.Siregar
Department of Astronomy and Bosscha Observatory
Bandung Institute of Technology, Bandung
Indonesia
e-mail :suryadi@as.itb.ac.id

## Abstract

*Since1984 asteroid study is conducted at Department of Astronomy, Bandung Institute of Technology. At present there are two main streams in our research, dynamical and physical study. Astronomers determine the orbital elements of more than 12000 asteroids, and the number is increasing. In the distribution of orbital elements of asteroids, there are several features such as Kirkwood gap, groups, and families. To know the dynamical evolution of asteroids, it is very important to study these features. Recently some small bodies of our solar system have been approached to the Earth, one of them was Toutatis. These objects are interested to study. Based on 2384 asteroids taken from NASA.File the phenomenon of orbital elements a, e, i, Ω, ω and Tisserand invariant, T of Near-Earth asteroids are briefly described*

## 1. Introduction

The aim of this study is continuation of previous long-term dynamical study of Amor-Apollo-Aten (collectively referred as AAA asteroids). Two main theories introduced to explain the origin of the asteroids in the main-belts. The oldest assumes the existence of a parent planet perturbing between Mars and Jupiter, which broken. Some of them ejected to libration points L4 and L5, called Trojan asteroids and others become the Earth-Crossing objects, well known as NEA (Near-Earth Asteroid) or NEC (Near-Earth Comet), (Siregar and Nakamura,1989). Nowadays, one theory largely accepted is that, asteroid are small condensations from the primitive solar nebula that were unable to form into a single body, undoubtedly because of the gravitational instability caused by the presence of Jupiter. These objects moving between Mars and Jupiter and collisions are happened frequently. A typical velocity is about 5 kilometers per second, when a small asteroid hits a larger one, it will make a crater in the larger body. If two objects are of comparable size, they will destroy each others, leaving hundreds or thousands of fragments. This collision destruction has occurred so often during the lifetime of the solar system, that practically all the asteroids we now see are fragments of the original parent bodies. Some may be found in unstable zone, like those of the Kirkwood gaps, they became the source of Apollo-Amor-Aten (AAA) asteroids. Study on the ecliptical motion and perihelion distribution resulted that asteroids occupy the ecliptic plane, implying that the selection of an exponential distribution function is reasonable. The maximum distance, $z_0$ to the ecliptic plane of Aten, Apollo and Amor represents an exponential distribution of, $f(z_0) = ab^{z0}$ . Here $z_0 \cong a \sin i$, and $|b|<1$ (Siregar,1991a) . It is estimated that there are about 1000 Apollos, 100 Atens and 1000-2000 Amors. Their mean life times are of the order of $10^7$ - $10^8$ years before they either are ejected from the Solar System, or collide with a planet (Siregar, 1991b). Simulation done by Yamada(1990) shows the Earth or Venus gravitational attraction had reduced comet's aphelion distances when the comets had passed close to the Earth or Venus and it has been found that there exist an analogy between the distribution of Jacobi constants of Amor-Apollo objects and those hypothetical comets that enter shallowly inside or graze the orbit of the Earth or Venus(Siregar,1993a). In additions it is interesting that so many of the AAA asteroids appear unusual, as compared with main-belts objects . The precise physical mechanism that might explain the differences remains a mystery. The present situation estimate that some 60% of the Earth-crossing (Apollo-type) objects supplied from some other reservoirs. At present the short period comets are the only identified reservoir that can fill that gap (Richard et al, 1992)

Study based on the view of Tisserand's invariant, $T = 1/a + 2[a\ (1-e^2)]^{1/2} \mathrm{Cos}\ i$ , where semi major axis, a, in unity of semi major axis of Jupiter, show that, T, of AAA objects spread in the range from 2 to 7.6 with strong concentration at $T \geq 3$, meanwhile comets have, $T< 3$ .There are 17 asteroids, consist of 7 Apollos and 10 Amors have a small value of Tisserand invariant. It is thus possible that asteroid could be of commentary origin. In this sense, they ought to be observed with greater attention. The other hand, there are 77 comets has a big value of Tisserand invariant, they often pass near Jupiter and their orbits forces, the comet's escaping gases give slight push to the comets, which slowly changes their orbit. These comets have a short period, P less than 8.6 years. It is also known that comet gradually decay. These appear to be real transitional case with regard to the comet-asteroid classifications objects, which are asteroidal in physical appearance, but they are cometary in dynamical behavior. These objects can be either extinct cometary nuclei or asteroid dynamically transferred from their usual stable orbit. By using 3rd order of Gauss equation some aspects to determine the orbit especially, stability and condition for good computation, has been discussed by Siregar (2001). In certain case some asteroids dynamically present as comets for examples near-earth asteroids, 1997 SE5, 1982 YA, 2002 RN38 and 2002 VY94. Partial part of comet trajectory can be determined by using Baker equation, therefore it is important to estimate the profile of orbit. Already well known that classical method to locate the real root of polynomial is algebra theorem of Descartes, detail of it is application see Siregar (1998). At the same time many of the cometary candidates have physical properties that are inconsistent with our current understanding of cometary nuclei such as albedos and (or) unusual spectra. (Siregar, 1995a, 1995b). Otherwise recently study by using least square method applied for 2384 Near-Earth asteroids suggest that distribution function of semi major axis of NEA are follow $F(x) = \alpha\ x^{\beta}\ e^{\gamma x}$. For degrees of freedom (d.o.f) = 31, we found that correlation coefficient, r = 0.95 with $\alpha = 7.84$, $\beta = 7.69$ and $\gamma = -5.16$. Mean while there are the evidence for a Near-Earth-Asteroid belt, composed of objects with semi major axis, $a \cong 1.4$ AU. The maximum of semi major axis is 17.96AU (1999 XS35) and the minimum of semi major axis a = 0.64 AU belong to 1999 KW4 (Siregar, 2003)

## 2. Regression Curve of the Orbital Elements

Empirical models of semi major axis, a, eccentricity, e and inclination, i
described such as, density function

$$F(x) = \alpha x^{\beta} e^{\gamma x} \qquad (1)$$

where

x = midpoint interval
F = frequency

The expression (1) analog to the linear equation;

$$Z = a_0 + a_1 X + a_2 Y \qquad (2)$$

In this case,

$Z = \mathrm{Log}\ F(x)$, $X = \mathrm{Ln}(x)$ and $Y = x$
$a_0 = \mathrm{Log}\ \alpha$ , $a_1 = \beta$ and $a_2 = \gamma$

Consider a sample $(X_1,Y_1,Z_1), \ldots, (X_n,Y_n,Z_n)$ of size n, and assume,
$S = \sum ( Z_i-( a_0 + a_1X_i + a_2Y_i))^2$

In the method of least square $a_0$, $a_1$, and $a_2$ are chosen such that S is minimum, the necessary condition for S to be minimum is;

$\partial S/\partial a_0 = 0$, $\partial S/\partial a_1 = 0$ and $\partial S/\partial a_2 = 0$ (3)

This yields the normal equation, which are given by;

$$\sum Z = na_0 + a_1\sum X + a_2\sum Y$$
$$\sum XZ = a_0\sum X + a_1\sum X^2 + a_2\sum XY \qquad (4)$$
$$\sum YZ = a_0\sum Y + a_1\sum XY + a_2\sum Y^2$$

The regression coefficient of (Z on X and Y), $\underline{a_0}$, $\underline{a_1}$, and $\underline{a_2}$ can be calculated by Cramer method. According to Spiegel(1982) if we define

$$\underline{Z_i} = \underline{a_0} + \underline{a_1}X_i + \underline{a_2}Y_i \qquad (5)$$

unexplained variation

$$\sigma_u = \sum ( Z_i - \underline{Z_i})^2 \qquad (6)$$

total variation

$$\sigma_T = \sum ( Z_i - Z_r)^2 \qquad (7)$$

$Z_r$ – average of $Z_i$

explained variation

$$\sigma_e = \sigma_T - \sigma_u \qquad (8)$$

then correlation coefficient is;

$$r = (\sigma_e / \sigma_T)^{1/2} \qquad (9)$$

Where the value of correlation coefficient is, $r \leq 1$

3. Results of Orbital Studies

Some figures of our recent study such as histogram of orbital elements, and plot of Tisserand invariant are presented in this section. The surface of zero velocity relative of asteroid, calculated according to the three-body problem. For more detail information see Danby (1989)

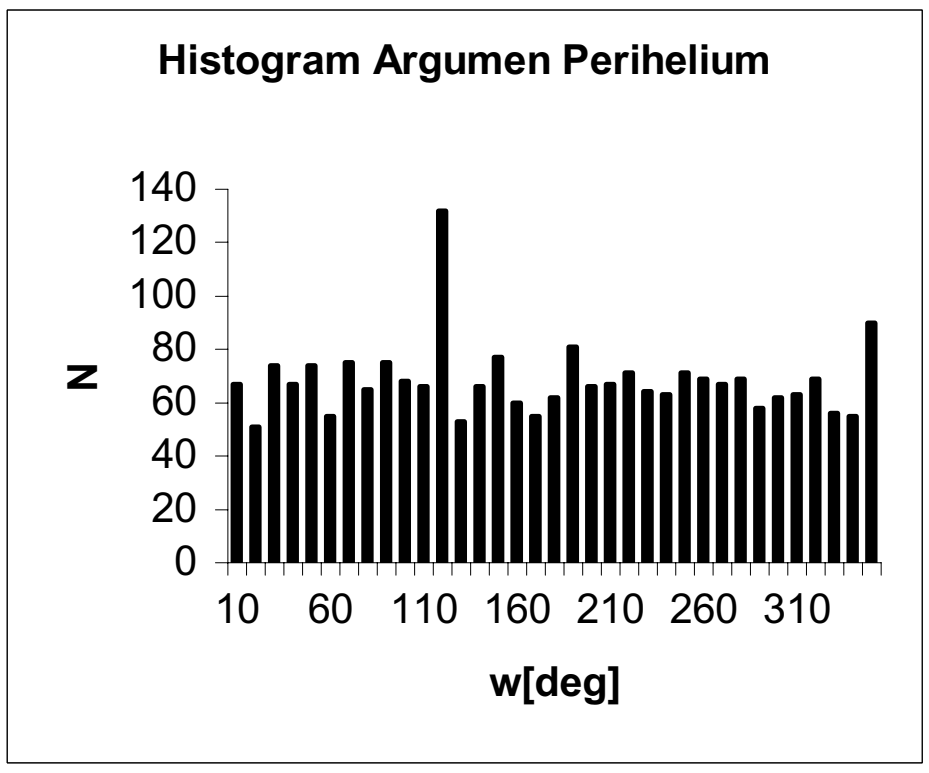

Fig 1.Histogram of Perihelion Argument of NEA. Total sample is 2383 asteroids NEA distribute evenly on ecliptic plane, ω maximum = $359^0.926$ (2001 BA40) ω minimum =$0^0.101$ (1999 BL33)

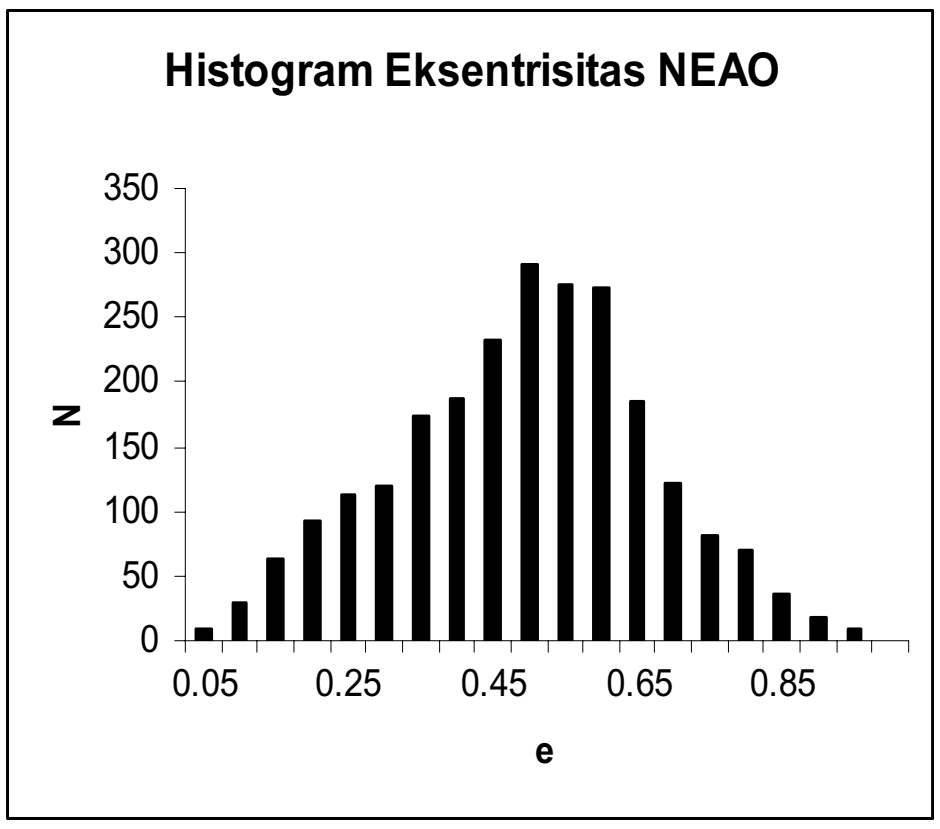

Fig. 2. Histogram of Eccentricity of NEA. Total sample is 2383 asteroids eccentricity maximum, e max= 0.956 (2002 PD43) Eccentricity minimum , e min = 0.013 (2002 AA29). Distribution function follows, $F(X)= \alpha X^{\beta}e^{\gamma X}$ with, $\alpha = 47.74$, $\beta = 2.73$ and $\gamma = -9.52$, correlation coefficient, r = 0.96, degree of freedom d.o.f = 19

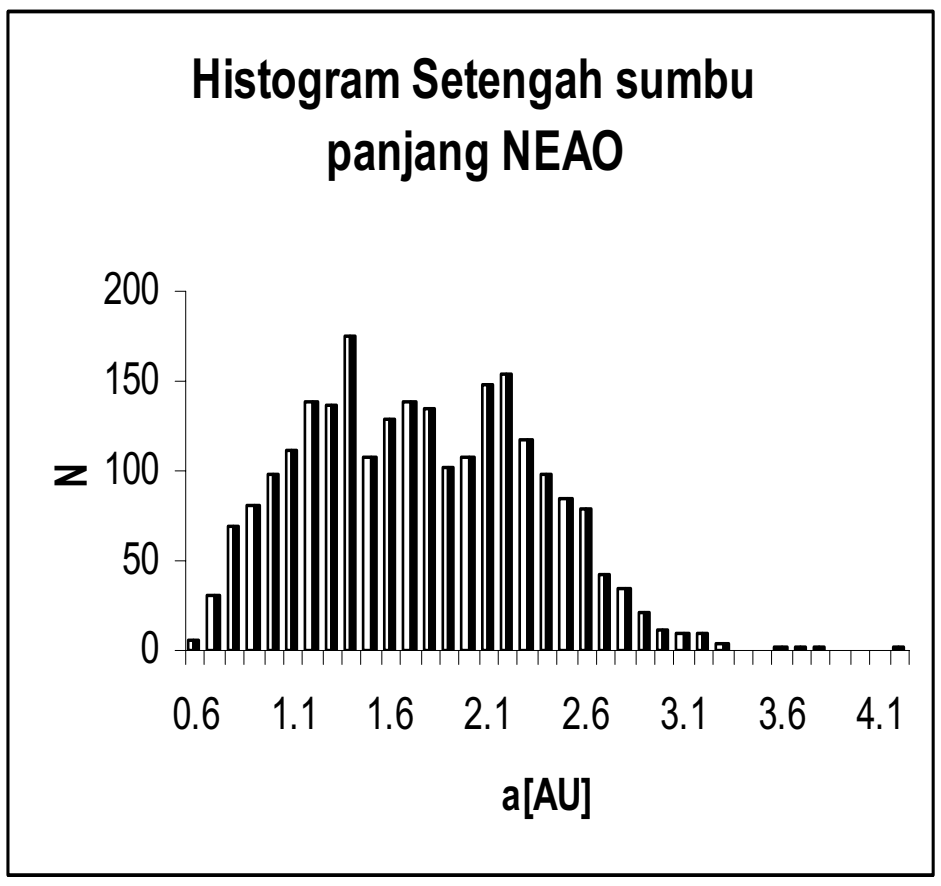

Fig.3 Histogram of Semi Major Axis of NEA, a maximum=17.976 AU(1999 XS35) and a minimum= 0.643 AU(1999 KW4). Total sample is 2382 NEA. Distibution function is, $F(X)=\alpha X^{\beta} e^{\gamma X}$ where, $\alpha = 7.84$, $\beta = 7.69$ and $\gamma = -5.16$, correlation coefficient is .r = 0.95. Degree of freedom.o.f = 31

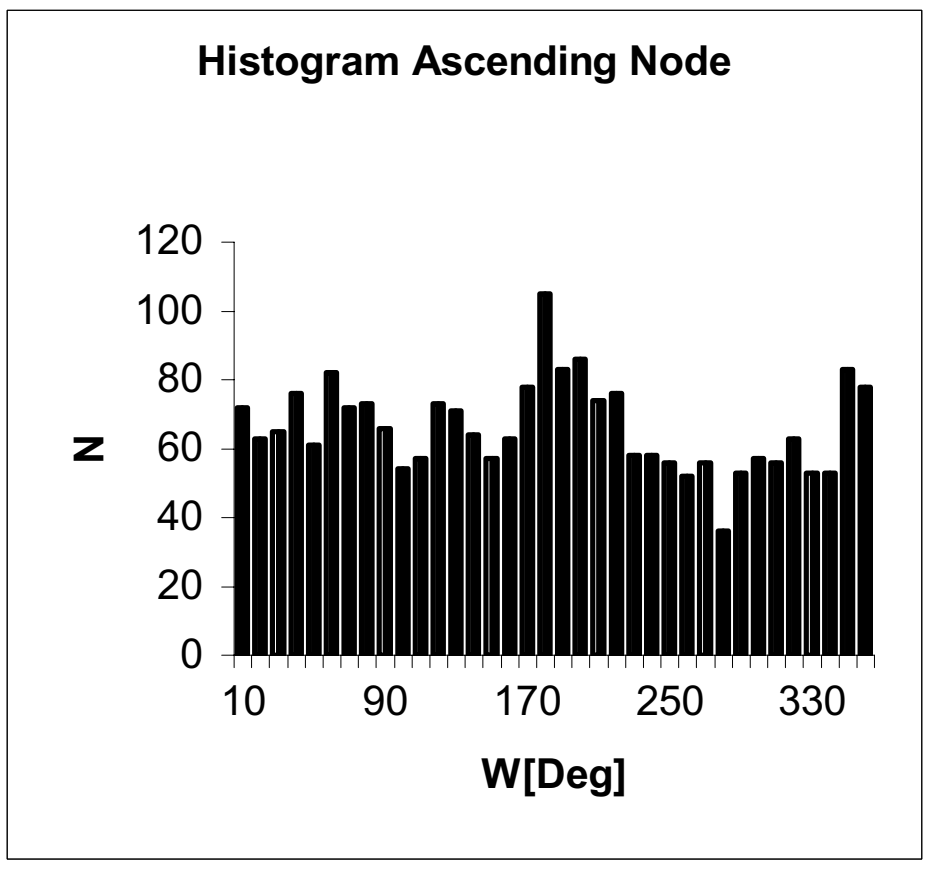

Fig. 5 Histogram of Ascending Node NEA. Largely concentrated on Autumnal Equinox. NEA distributed evenly on ecliptic plane. Ω maximum = $359^{0}.919$ (2000 EY106). Ω minimum $=0^{0}.008$ (11885 1990 SS). Total sample is 2383 NEA

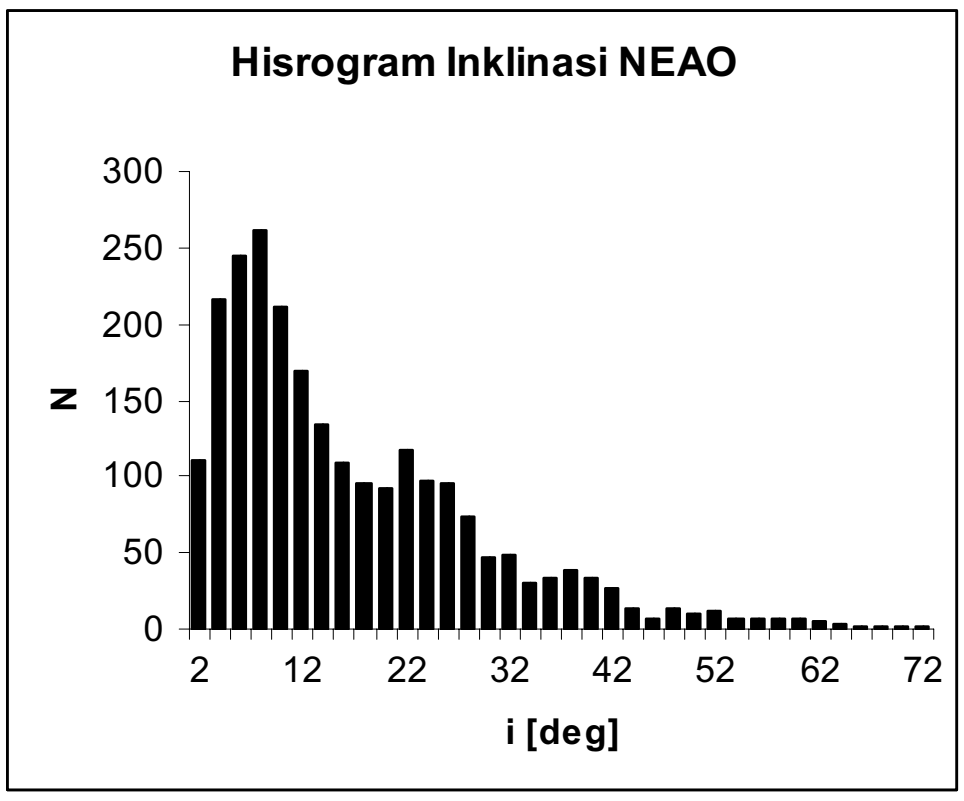

Fig.4.Histogram of inclination of NEA. Total sample is 2382 i maximum= $72^{0}.143$ (2001 AU43), i minimum= $0^{0}.0.110$ (2000 SG344) Distribution function $F(X)=\alpha X^{\beta} e^{\gamma X}$ $\alpha=0.05, \beta=0.66$, and $\gamma = -0.11$. Correlation coefficient, r = 0.98, Degree of freedom , d.o.f = 35

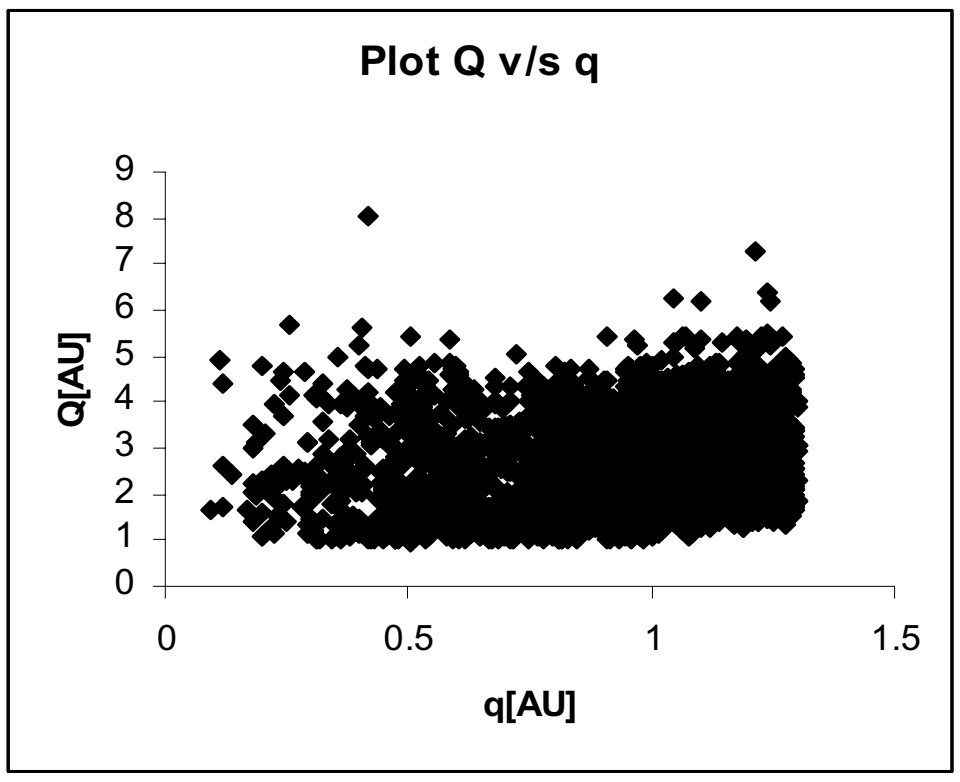

Fig 6. Plot Aphelion v/s Perihelion 0f 2382 NEA Maximum of perihelion and aphelion respectively are q=1.300 AU(2002 PG43) Q=34.999 AU(1999 XS35) Minimum of perihelion and aphelion respectively are q=0.092 AU(2000 BD19) Q=0.98 AU(2003 CP20)

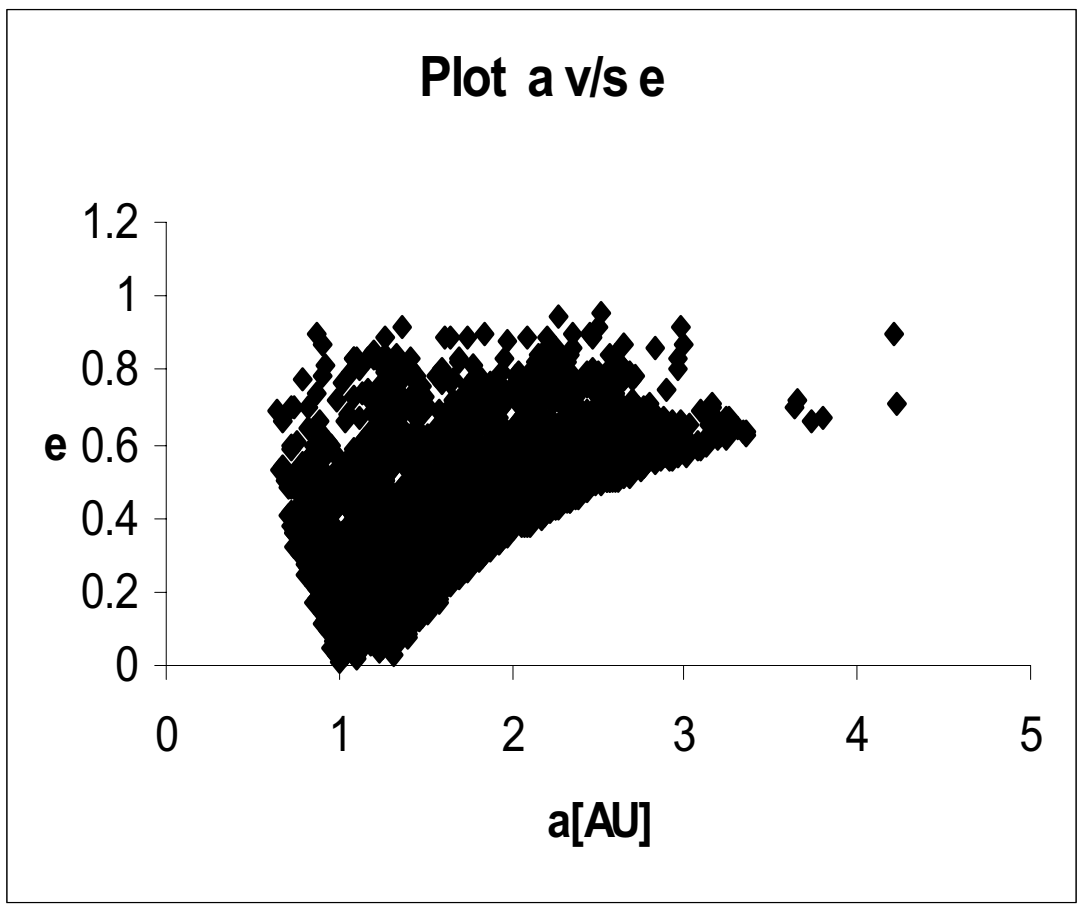

Fig. 7. Plot Eccentricity v/s Semi Major Axis of NEA. Maximum semi major axis and eccentricity respectively are, a = 17.976 AU (1999 XS35), and eccentricity e = 0.956 ( 2002 PD43). Minimum semi major axis and eccentricity respectively are, a = 0.642 AU(1999 KW4) and e = 0.013 (2002 AA29). Total sample are 2382 asteroids

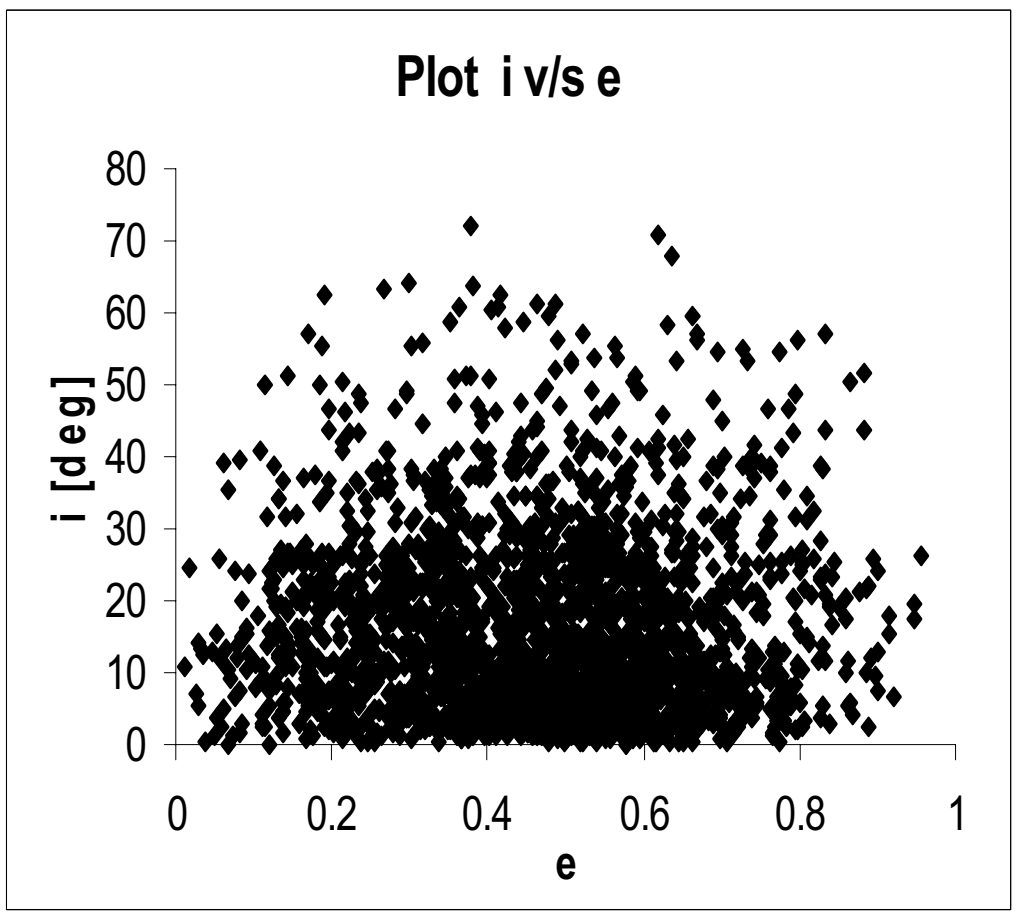

Fig. 9 Plot of inclination v/s eccentricity of 2382 NEA. Maximum of inclination and eccentricity respectively are, $i=72^0.143$ (2001 AU43), and e= 0.956 (2002 PD43).Minimum of inclination and eccentricity respectively are, $i=0^0.110$ (2000 SG344) e=0.013 ( 2002 AA29)

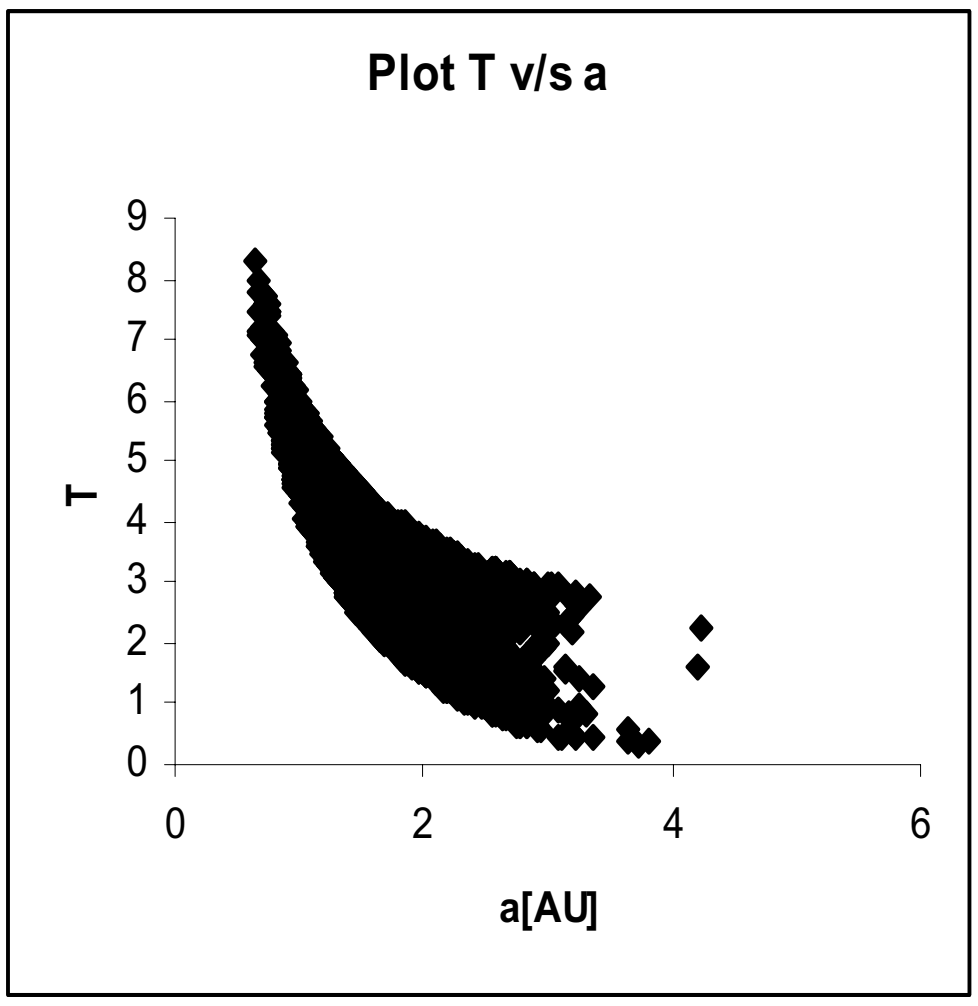

Fig.8. Plot of Tisserand Invariant v/s Semi Major Axis NEA

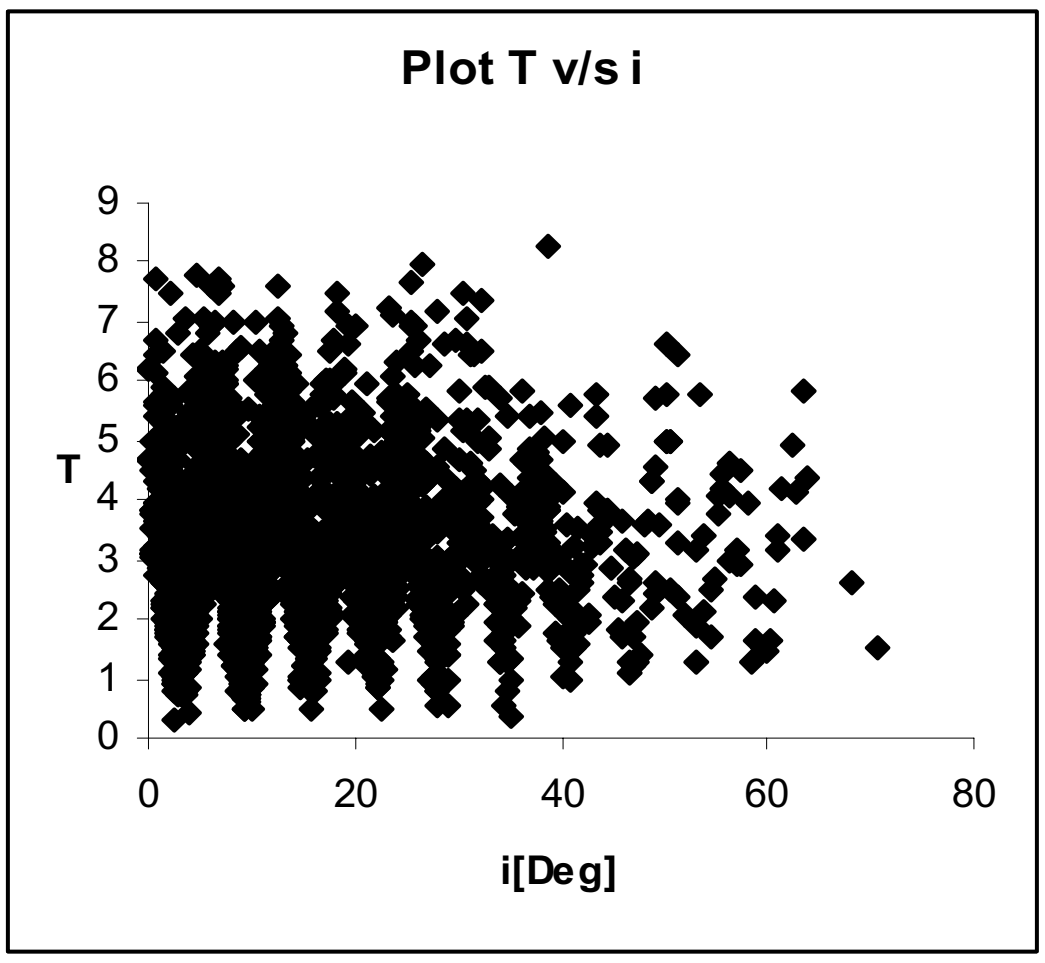

Fig.10. Plot of Tisserand Invariant v/s Inclination of NEA

The relative zero velocity surfaces

The general problem of the motion of the three bodies (assumed to be point masses), subject only to their mutual gravitational attractions has not been solved, although many particular solution have been found. Two bodies of finite mass revolve around one another in circular orbits (Sun and Jupiter), and a third body of infinitesimal mass (asteroid) moves in their field., this situation is approximately realized in many instances in the solar system. Figures 11, 12 and 13 show three aspects of the surface;

$$x^2 + y^2 + \frac{2(1-\mu)}{r_1^2} + \frac{2\mu}{r_2^2} = T \qquad (10)$$

where the distance in unity distance primary to secondary. Total masses is 1 and $\mu$ secondary mass. The distance asteroid to primary is

$$r_1^2 = (x-\mu)^2 + y^2 \qquad (11)$$

and the distance to secondary mass is

$$r_2^2 = (x-1+\mu)^2 + y^2 \qquad (12)$$

T is a constant where the velocity of third body equal zero well known as Jacobi's integral or Tisserand invariant. Although there are many asteroids in our solar system, the size of an individual asteroid id quite small. The largest asteroid known is Ceres, which has a diameter of about 910 km. The others are much smaller, so the masses of the asteroids are negligibly small in comparison with those of the planets. There fore the mass of asteroid do not consider

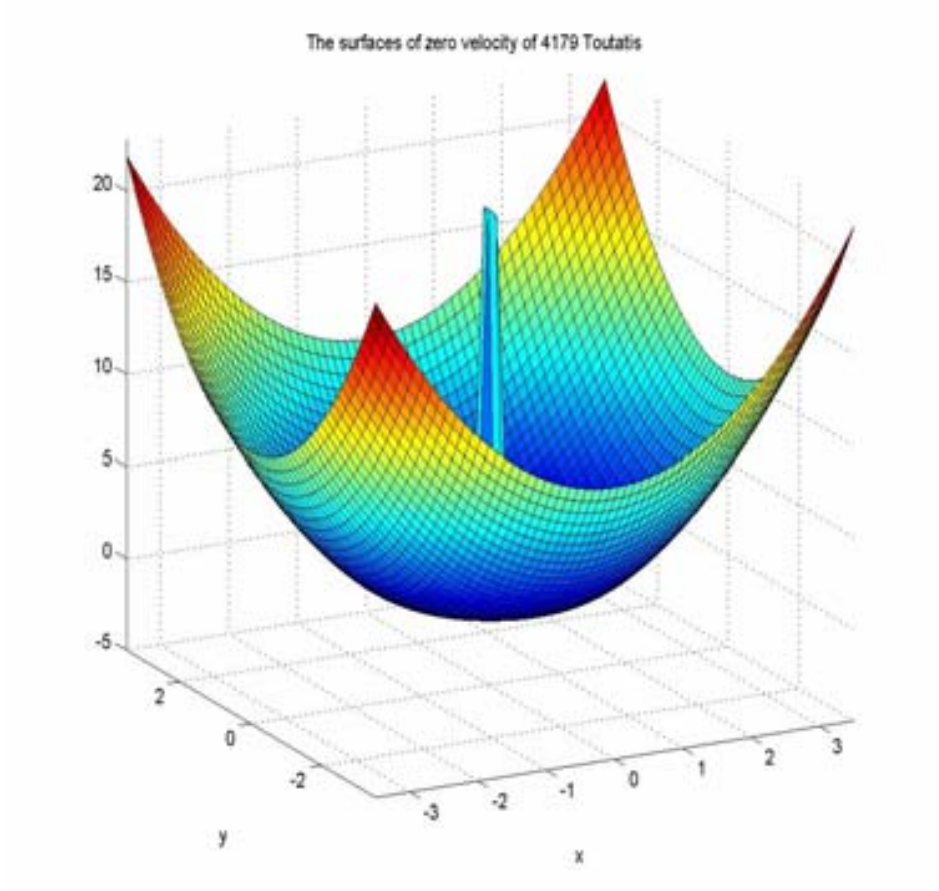

Fig.11.The surfaces of zero velocity of near earth asteroid 4179 Toutatis, Tisserand invariant T= 3.0297 and $\mu$ = 0.001

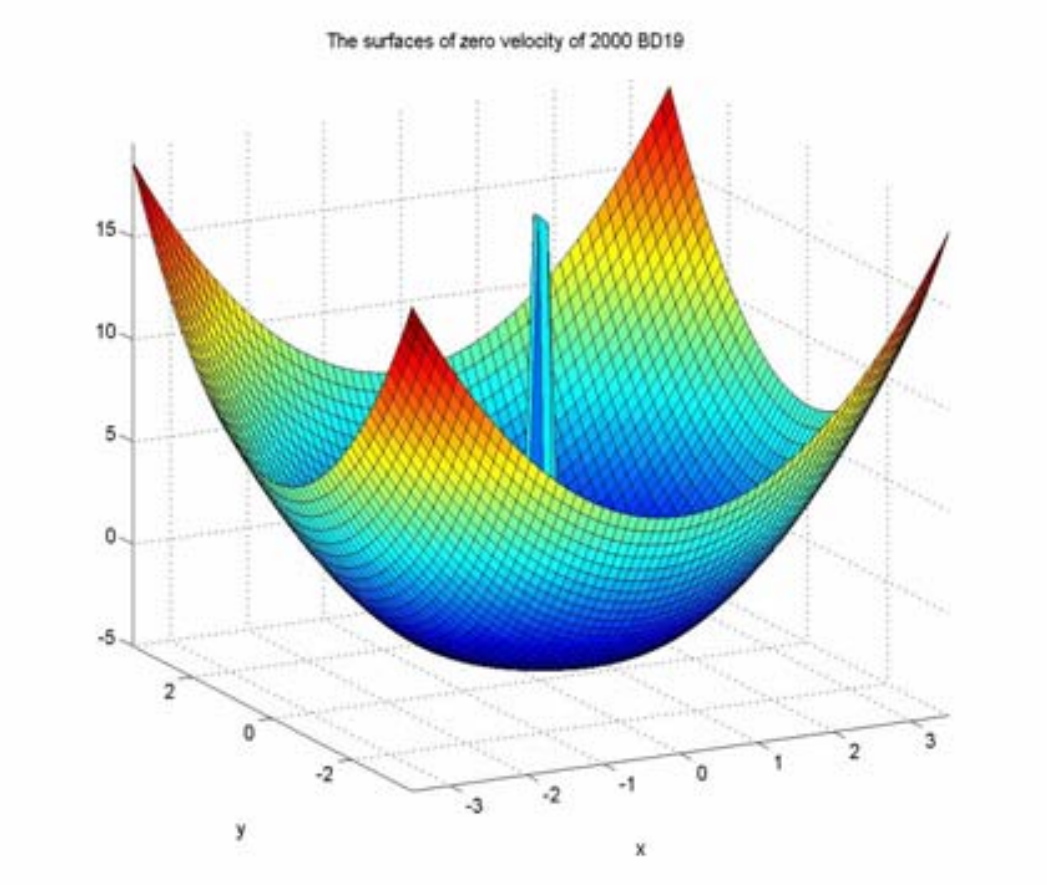

Fig. 12. The surfaces of zero velocity of 2000 BD19, Tisserand Invariant T=6.247 and $\mu$ = 0.001.

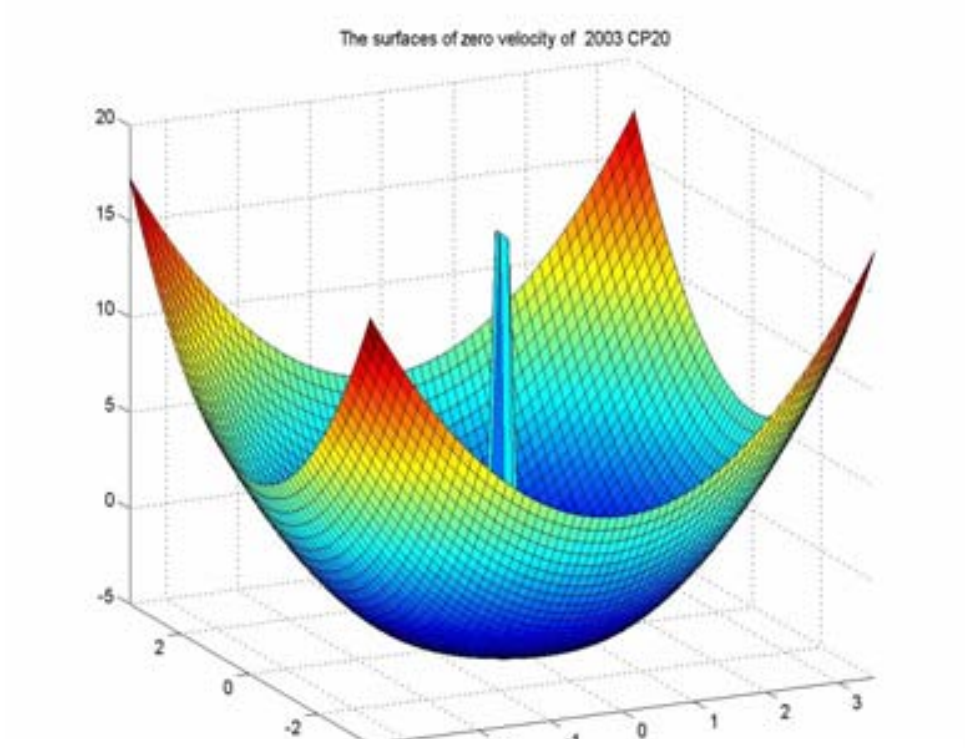

Fig. 13. The surfaces of zero velocity of 2003 CP20. Tisserand invariant T=7.649, $\mu = 0.001$

## Discussion

These work let some remarkable questions. Table 1. showed selected objects from our study, some NEA presents the Tisserand invariant less than 1, dynamically they have comet's behavior, and physically they are asteroids.

Table 1. Comet like asteroids

| Name | a[AU] | e | i[Deg] | T |
|---|---|---|---|---|
| 1997 SE5 | 3.730 | 0.666 | 2.608 | 0.31 |
| 1982 YA | 3.642 | 0.698 | 35.056 | 0.37 |
| 2002 N38 | 3.801 | 0.675 | 3.850 | 0.41 |
| 2002 Y94 | 3.237 | 0.659 | 9.165 | 0.46 |

Are they extinct or dormant comets ?.